# Intrusion Detection in Computer Systems by Using Artificial Neural Networks with Deep Learning Approaches

Sergio Hidalgo-Espinoza, Kevin Chamorro-Cupuerán and
Oscar Chang-Tortolero

School of Mathematics and Computer Science,
University of Yachay Tech, Ecuador

## ABSTRACT

*Intrusion detection into computer networks has become one of the most important issues in cybersecurity. Attackers keep on researching and coding to discover new vulnerabilities to penetrate information security system. In consequence computer systems must be daily upgraded using up-to-date techniques to keep hackers at bay. This paper focuses on the design and implementation of an intrusion detection system based on Deep Learning architectures. As a first step, a shallow network is trained with labelled log-in [into a computer network] data taken from the Dataset CICIDS2017. The internal behaviour of this network is carefully tracked and tuned by using plotting and exploring codes until it reaches a functional peak in intrusion prediction accuracy. As a second step, an autoencoder, trained with big unlabelled data, is used as a middle processor which feeds compressed information and abstract representation to the original shallow network. It is proven that the resultant deep architecture has a better performance than any version of the shallow network alone. The resultant functional code scripts, written in MATLAB, represent a re-trainable system which has been proved using real data, producing good precision and fast response.*

## 1. INTRODUCTION

Information security is, nowadays, one of the most important topics in computer science. This is due to the giant internet-connected networks and devices that increase day to day, making hacking activities to increase in the same proportion. The fight against these never end activities is complex because there exist lot of reasons why an information system becomes an attractive target. These goes from activism (called 'hacktivism'in this environment), people eager to show their hacking abilities or just for fun. Also, everyday hackers develop new abilities and techniques to infringe security of information systems [1]. Therefore, it is necessary to constantly design new tools that fight against malicious activities. A fertile alternative to develop hacking protecting systems are Artificial Neural Networks.

Artificial neural networks (ANN) are data structures dedicated to get results to non-deterministic problems, in a different approach to another conventional process. Important areas where ANNs are actively used are data classification, mapping, prediction and clustering [2], for which it is necessary to have a compilation of data (the dataset) to get the desired results. In general, there exists 3 phases to execute an ANN: training, validation and testing. Training is the phase in which weights are modified to get the optimal level of learning. This phase could be performed using several modifications in the components of the network to get varied weights to be tried and obtain the best weights which will return an optimal classification. This phase uses the major amount of data, it is recommended to use 80% of data approximately. Validation phase is used to test the different weights obtained in training phase to choose the best ones and sometimes to get weights more tuned than before. Finally, testing phase is the fireproof to the network. At this step, the error must be minimal and impossible to reduce. In case the network does not return convincing results, it is necessary to check and modify the parameters of the network. Usually the remaining 20% of data is distributed to perform validation and testing phases.

ANNs are systems based in the human brain functionality [3]: an external signal comes to the system as input data (external senses of the human body), this information is carried to the core of the neural network (human brain) to be interpreted by *synapses* [4] between artificial neurons (human brain neurons). Finally, the result is showed as output data (reactions in the human body). Generally, the results of an ANN are led by a series of calculations performed by steps into several iterations or epochs [5]. Some ANNs may be more advanced and solve problems with better results by the use of mixed ANNs into a nested one, these particular ANN's are called Deep Learning[6][7][8].

Deep Learning Architectures do not have a strict definition (like ANNs), it means does not exist a number of neurons, layers or models to work with these tools. They are just another kind of data processing structures which apply high levels of abstraction [6], using methods to analyse and get information from data deeper (hence its name) than other ANNs, turning Deep Learning architectures on "black boxes" and allows to get better results which means higher percentage of patterns classification.

As a result of applying these concepts, it has been designed and built a Deep Learning structure to process logins into a computer network, classifying them as hacking attacks or normal activity, obtaining high levels of accuracy.

## 2. RELATED WORKS

The importance of this work lies in the fact of create an Intrusion Detection System prepared to learn and improve its accuracy by using advanced techniques, trained and tested with data updated to current times which could provide a level of adaptationready to detect any attempted intrusion. To achieve this objective, some researches in the same field have been studied, giving a clear idea about the direction of this work.

### 2.1. Y. Liu, S. Liu and X. Zhao (2017)

Liu [9] and his team designed an intelligent system applied to detect intrusion using another kind of Deep Learning architecture: Convolutional Neural Network (CNN) and other different techniques, as factors of comparison, which are not necessarily related to ANNs. They used the 10% of a famous database of intrusion attacks: the KDD Cup 1999 assembled by DARPA consisting of more than one million of patterns and 41 parameters to each pattern, applying 22 types different hacking attacks. After processing these data into the network, they obtained 99.7% of detection rate (capacity to perform a correct classification) using the CNN which was the better tool of that research. The problem of these results is, obviously, the age of the data, the world and computers do not work equal than the last century anymore. Testing this network with updated data will carry important changes and configurations into the same network, because new data will have new parameters to be processed. Also, this dataset has some problems, for which it is necessary perform pre-processing algorithms to clear the dataset, some of them are studied by McHugh [10].

## 2.2. Biswas (2018)

Biswas [11] performed a similar research than Liu, using different techniques for classification, such as Nave Bayes, Support Vector Machine, Decision Tree, Neural Network (no specified) and k-nearest neighbour algorithm (k-NN); in the other hand they are used feature selectors algorithms to take relevant characteristics of data, such as Correlation based Feature Selection method (CFS), Principal Component Analysis (PCA), Information Gain Ratio (IGR) and Minimum Redundancy Maximum Relevance. The difference of this research is the use of a modified version of KDD 99 dataset, the NSL-KDD. This improved version is based on the problems studied in the original dataset, detailed by Tavallaee [12] in 2009. The NSL-KDD consists of 100000 patterns of activity and more than 40 parameters, reduced to 10000 in order to avoid problems with computational cost.The combination of k-NN with IGR gives best results of experiments, reaching a 99.07% of accuracy. The problem is the same than before, data was created in 1999 which does not have compatible metrics with current data, despite it was filtered and improved.

## 2.3. Vinayakumar, Alazab, Soman, Poornachandran, Al-Nemrat and Venkatraman (2019)

In this work, developed by Vinayakumaret al. [8], it is used a Multi-layer Perceptron as processing architecture, the Feed Forward algorithm and several activation functions such as sigmoid, tangent,softmax and ReLU; they treat the compilation of these characteristics as a Deep Learning architecture. Their results are divided into a lot of tests with different datasets and different configurations of the network, obtaining high percentages of classification overcoming the 90% in almost all the experiments.

## 3. THE DATASET

The dataset is one of the main components for the training of the network. It consists in a lot of patterns previously generated by experiments or observation; each pattern has many parameters which characterize it. Each one of the patterns may be as random as the experiments be, we do not know or could predict its behaviour just its nature. The inputs are presented as a CSV (Comma Separated Values) file in which each pattern is a new entry or row and each parameter is a new column. Each row must have all columns filled, in case there exist any column empty, it is necessary to have a default value to fill it, but it is better to remove that row.

## 3.1. CICIDS2017 Dataset

For this case, it is used the **Intrusion Detection Evaluation Dataset (CICIDS2017)** developed in 2017 by the **Canadian Institute for Cybersecurity (CIC)**[13] in laboratories dedicated to information security. The idea to build this dataset is performing a massive amount of different controlled hacking attacks from different parts of the world to dedicated servers which measure and register the chosen parameters to be saved as files. Finally, the result is thousands of patterns (hacking attacks or normal activity), which are the inputs patterns for this work, with 78 characterizing parameters and 1 target parameter (seen as CP and TP in Figure 2, respectively) which is the result to reach. Some important details of this dataset are studied by [14].

## 3.2. Data Normalization

Before data patterns could be processed by any ANN they must be processed by an algorithm of normalization, this is a process in which all patterns' values are placed in the same range, being [0-1] the most used. Normalization is important because major of data values are produced naturally in random ranges, it makes any ANN increase the effort to achieve optimal results or decrease the quality of them. Therefore, this process optimizes the capacity of the network to get the optimal classification for each pattern.

Figure 1 indicates how data does not change its behaviour after normalization but it takes a different shape in order to be easy to process by the network. It is just adjusting the numerical range of data between [0-1] instead of its natural range using the next equation for each one of the characterizing parameters (columns of the dataset):

$$n_i = \frac{x_i - \min(x)}{\max(x) - \min(x)} \qquad (1)$$

where:

$x$: the characterizing parameter in the dataset,
$i$: the number of each pattern,
$n$: the resultant column after normalization.

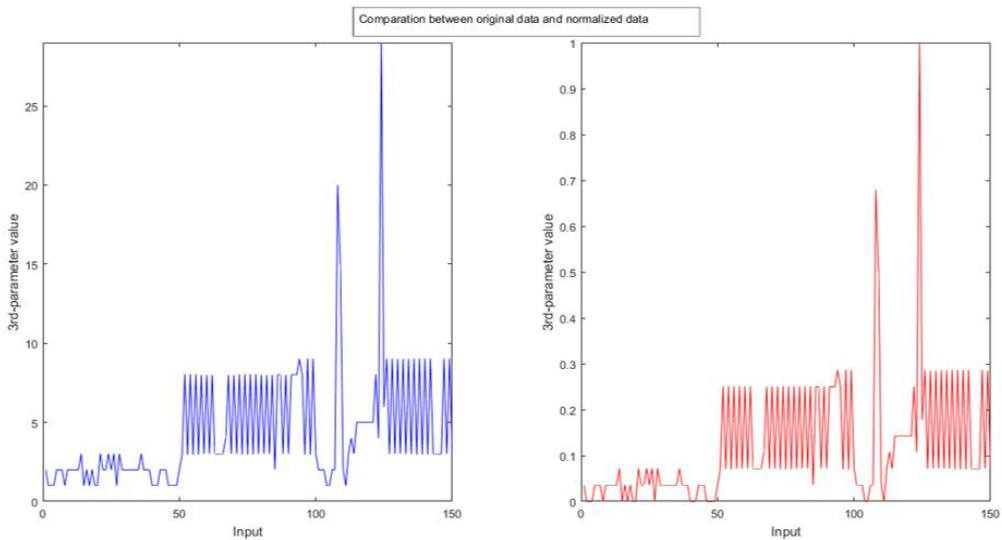

Figure 1. Data normalization for a single parameter using a 30-patterns dataset. Blue: Original data. Red: Normalized data between [0-1].

It is not necessary to normalize the target pattern because it is just a boolean output: 1 if the correspondent pattern is an attack, and 0 in the opposite case.

## 4. DEEP LEARNING ARCHITECTURE

In order to demonstrate the utility of this Deep Learning model, the experiments of this work are performed in two different ANN architectures: **a shallow network** which consists of a ***Back-propagation Algorithm*** (**BPA**) only, and a **deep network** formed by two single ANN architectures: a two-layered **auto encoder** and the BPA mentioned before. In this case, the autoencoder is a complement architecture dedicated to turn pure data into compress data easy to process by the BPA. Autoencoders usually have two parts: encoder and decoder, the encoder takes data and extract recondite information which may not be visible to the ANN. This process is performed by a set of equations and calculations (because autoencoders are a kind ANN such as) which leave as result the weights of the encoding and set of compressed data (the encoded data). In the other hand, the decoder takes encoded data and its respective weights to obtain the input data again. The relevance of the encoded data to work along the BP algorithm depends on the accurate when data is decoded back, it means the decoding error must be minimal to have useful encoded data.

The BPA used into the two architectures will be almost the same. For the first one (the shallow network), it has 3 layers: the **input layer**, the **hidden layer** and the **output layer** as showed in Figure 2. Every layer is formed by a number of neurons previously selected, in this work it was established the next distribution: 78 layers for the input layer, 11 for the hidden layer and one for the output layer which is compared with the target parameter to get the error between them. Finally, the result is obtained from this error after executing the number of proposed iterations.

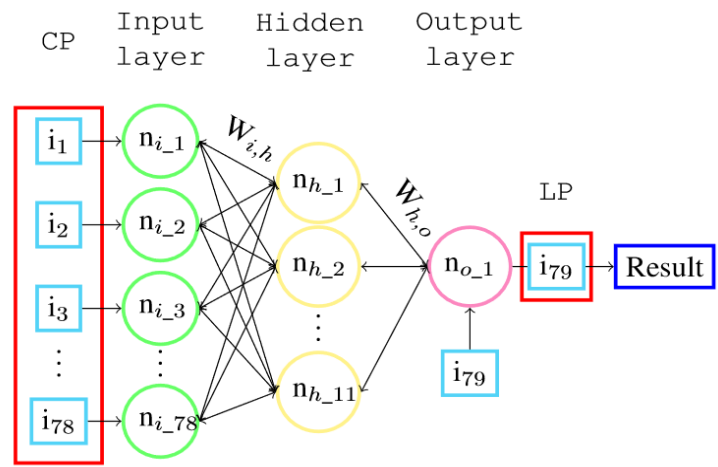

Figure 2. First ANN architecture: Shallow backpropagation topology. CP: Characterizing parameters. W: Weights. TP: Target parameter.

One of the most important part of the BPA are the **weights (W)**, which are data matrices whose values are calibrated in the process of ANN's learning. This calibration corresponds to "recording" the learning of ANN between pairs of layers (each layer with its adjacent layer) in each iteration of the ANN. The level of recording the learning in each iteration depends on a factor called the **learning rate (LR)** which determines the level of remembering what was learned in each iteration. For this research, it is used a value of 0.1 as LR and weights are initialized randomly between the selected range, it is achieved just using a random function:

$$W = \text{rand}([-1, 1])$$

Inside the tangled process of calibrating the weights between layers, there exist a function dedicated to calculate the performance of every neuron into the network layers, this performance is calculated from the output of the neuron, called the **synaptic potential** (**SP**)evaluated into an equation called **transfer function**. The output of this function is the real output of the neuron, consisting in an equation that could be selected depending on the applications of the network, even it could be designed by each developer, but in this research is used the so famous**logistic or sigmoidal function**, represented by:

$$g(h) = \frac{1}{1 + e^{-\beta h}}$$

(2)

where:

$g$: the transfer function,
$h$: the synaptic potential of the neuron,
$\beta$: is just a regulator of the SP variability, in this case it is 1, therefore it does not affect the results.

## 5. DEEP LEARNING IN ACTION

### 5.1. Feeding the network with data

The input patterns with their 78 characterizing parameters get into the input layer, one parameter per neuron (it means the number of neurons in the input layer is the same than the number of the characterizing parameters). Each pattern is provided only one time by iteration. In each iteration, every pattern is taken randomly, it allows cleaning the network from any **cyclic learning**, it is not an official term but sounds good to express a repetitive learning which could calibrate the weights in a monotonous way obtaining as result constant values for the matrix weights.

### 5.2. Processing data through the shallow network

Once into the input layer, input patterns are processed along with the input-hidden weights using the transfer function, these results will give rise to the hidden layer, which in turn are processed together the hidden-output weights, using the transfer function too. Then, when the process reaches the output layer it comes back from the output layer to the input layer optimizing the value of the weight's matrices. Then, the process goes back to the output layer, returning the output for this pattern in form of a probability to be or not a hacking attack. If this probability is greater than 50%, the pattern is considered an attack, in the opposite case it is not an attack (1 or 0). Finally, this result is compared with the target parameter to get the error using the squared error function:

$$E_t = \frac{1}{N} \sum_{i=1}^{N} (Y_i - Z_i)^2$$

(3)

where:

$E_i$: the total error of the network,
$N$: the number of patterns,
$Y$: the output of the network,
$Z$: the target (label parameter in input patterns).

This process is repeated by every pattern in each iteration, performing a reiterative process the number of times (iterations) previously selected or when a minimum error is reached.

### 5.3. Processing data through the Deep Learning network

The process in this architecture is almost the same at the last one. The difference is a previous step before data introduction into the input layer. This step is the autoencoder, which replace the old input layer. Now, data is introduced directly into the autoencoder which will extract information digging deeper between the great amount of introduced data. Input data is introduced only once into the autoencoder, it means the outputs of the autoencoder will be the new inputs of the shallow network, whose number of neurons in the input layer has been decreased, because it does not take all inputs anymore but it takes the extracted data resultant from the encoder. In this network, the number of the input layer has been reduced to 19 and the other layers keep the initial shape. A nice visualization of the new and deep network is showed in Figure 3. The BP algorithm of the deep network works exactly the same as the shallow network.

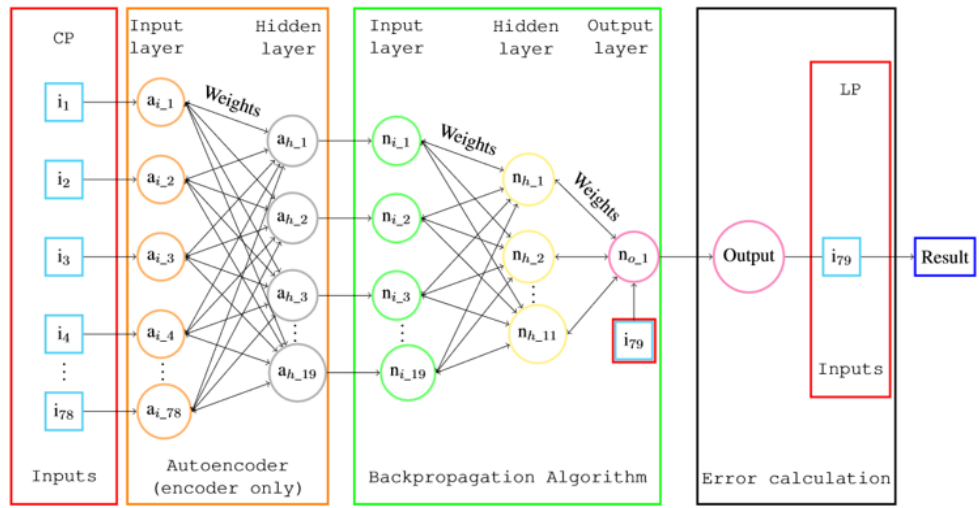

Figure 3. Ultimate visualization of the Deep Learning Network. Inputs have the same shape than before. Encoder is added. The shallow network (BPA) reduces its number of neurons.

### 5.4. Obtaining the final results

The most important results of this research lie in the fact of detecting hacking attacks in each one of the logs (input patterns) performed into processed data. In order to obtain the accuracy of the network, in both cases detecting an attack or discarding it, the testing phase results will pass new accuracy tests: **sensibility** and **specificity**, calculated by equations (4) and (5) respectively. Sensibility measures the proficiency to detect an attack when it is truly happening (true positives TP), a fail into the network will trigger a false positive (FP). Meanwhile, specificity takes care of cases which are not an attack but are normal activity (true negatives TN), a misclassification will lead a false negative (FN) which are more dangerous than FP because the non-detection of a real attack avoids a response against the attack.

$$Sensibility = \frac{TP}{TP + FN} \qquad (4)$$

$$Specificity = \frac{TN}{TN + FP} \qquad (5)$$

## 6. RESULTS

In order to show a good organization and prove the reliability and accuracy of the ultimate Deep Learning network, the results are shown in the same sequence than they have been performed: first using the shallow network and finally using the Deep Learning network. Every test use 1000 iterations for the autoencoder (do not confuse with the ANN's iterations), except for the last one, which uses 3500 iterations.

### 6.1. The Shallow Network

#### 6.1.1. 150 inputs

At 1000 iterations/epochs it is reached an error of **0.0434** in the last iteration. It looks good because it means a performance of **95.66%** but 150 [inputs] is not a trustworthy number for an optimal training and Figure 4a is not exactly the objective to achieve in this research because these horrible fluctuations between iterations 50 and 250.

#### 6.1.2. 6000 inputs

Let's use more data to work with the ANN. The first training with this dataset consists of 300 iterations resulting an error of **0.0961** and performance of **90.39%**, it does not look good anymore. Also, these peaks in Figure 4b shows a non-satisfactory training either, it evidences a low stability in the network given the extensive error variation between iterations which avoids a regular training.

Now, running 1000 iterations it is obtained an error of **0.0873** and a performance of 91.27\% but Figure 4c shows the same behaviour than before which is not reliable: the poor stability of the network.

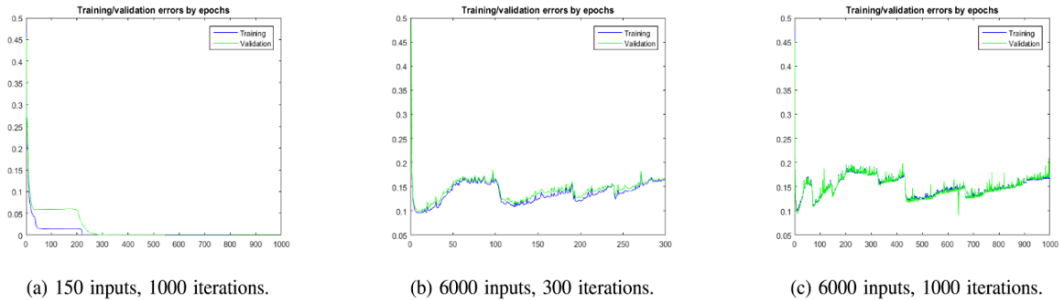

(a) 150 inputs, 1000 iterations.    (b) 6000 inputs, 300 iterations.    (c) 6000 inputs, 1000 iterations.

Figure 4. Training and validation results of the Deep Learning network.

### 6.2. The Deep Learning Network

#### 6.2.1. 300 iterations

As a confidence prove to the final architecture, it has been selected directly 6000 inputs, number to be used in all experiments for this network. Using only 300 iterations it is reached an error of **0.0084** which results in a performance of **99.16%**. It is a great percentage of accuracy and Figure 5a shows a behaviour better than any of the shallow network training.

### 6.2.2.  500 and 1000 iterations

Increasing the iterations to 500 results have improve, as expected, obtaining an error of **0.0068** corresponding to **99.32%** of performance, numbers which are obtained using 1000 iterations too, which demonstrate a stable behaviour in the network. The plotting of error by iterations does not change as seen in Figure 5b and5c.

### 6.2.3.  5000 iterations

It is just a checking to know whether the network will response in an unusual way given the large amount of iterations. The result is a performance better than any other one seen before in this research, reaching an error of **0.0060** equivalent to **99.40%** of performance. Obviously, the visualization is better too (Figure 5d). It is the better performance of the network to all experiments executed.

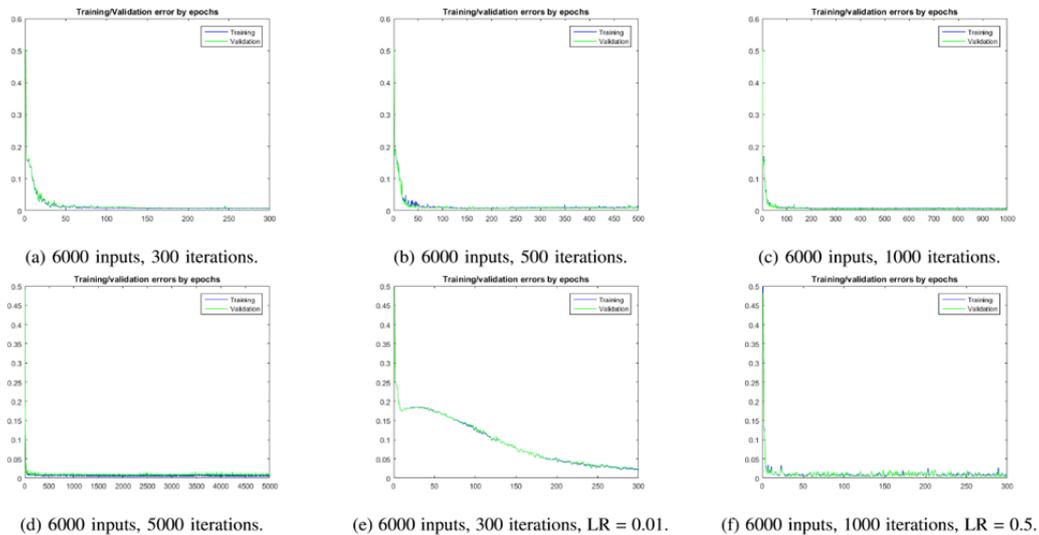

(a) 6000 inputs, 300 iterations.        (b) 6000 inputs, 500 iterations.        (c) 6000 inputs, 1000 iterations.

(d) 6000 inputs, 5000 iterations.   (e) 6000 inputs, 300 iterations, LR = 0.01.   (f) 6000 inputs, 1000 iterations, LR = 0.5.

Figure 5. Training and validation results of the Deep Learning network.

## 6.3. Other modifications in the Deep Learning Architecture

### 6.3.1. LR = 0.01

Decreasing ten times the LR selected to the last experiments gives as result a singular behaviour in the network, in which error is minimized in the last iterations. In the first iterations it is visualized a high error in respect to the last training. Also, it is appreciated a rise of the error which are not a good signal for the purposes of this work. Figure 5e shows all these details clearly. The error obtained **0.0165** and a performance of **98.35%**.

### 6.3.2. LR = 0.5

This experiment is the opposite case of which was made before: increasing the learning rate to 0.5. In this case the behaviour is similar to the experiments with LR=0.1 with an error of **0.0090** and a performance of **99.10%** but it is notorious the existent fluctuations in the error (Figure 5f) which is derived in low stability. In any case, the LR=0.1 with 5000 iterations is still better given its performance and stability. A summarize of all results is visualized in Table 1.

Table 1. A summary of results from different architectures of the ANN.

| N. | # inputs | Iterations | LR | Autoencoder | Error | Performance (%) | Stable |
|----|----------|-----------|------|-------------|--------|-----------------|--------|
| 1 | 150 | 1000 | 0.1 | No | 0.0434 | 95.66 | No |
| 2 | 6000 | 300 | 0.1 | No | 0.0961 | 90.39 | No |
| 3 | 6000 | 1000 | 0.1 | No | 0.0873 | 91.27 | No |
| 4 | 6000 | 300 | 0.1 | Yes | 0.0084 | 99.16 | Yes |
| 5 | 6000 | 500 | 0.1 | Yes | 0.0068 | 99.32 | Yes |
| 6 | 6000 | 1000 | 0.1 | Yes | 0.0068 | 99.32 | Yes |
| 7 | 6000 | 5000 | 0.1 | Yes | 0.0060 | 99.40 | Yes |
| 8 | 6000 | 300 | 0.01 | Yes | 0.0165 | 98.35 | Yes |
| 9 | 6000 | 300 | 0.5 | Yes | 0.0090 | 99.10 | No |

## 6.4. False positives and false negatives

Using the best results of Table 1, the experiment number 7: 2000 new patterns totally unknown to the network and equations (4) and (5) it is obtained Table 2 which shows the accuracy of the network, it means, the capacity of the network to label a pattern as hacking or normal activity.

Table 2. FP and FN results: using the best architecture.

| Total patterns | Well-classified | False positive |
|----------------|-----------------|----------------|
| 2000 | 1870 | 130 |
| **True Positive** | **True Negative** | **False Negatives** |
| 1000 | 870 | 0 |
| **Sensitivity** | **Specificity** | **Accuracy** |
| 1 | 0.87 | 93.5% |

An ultimate experiment is performed, changing the number of iterations for the autoencoder to 3500, the network performs its respective training and validations steps and results looks so much better (Table 3).

Table 3. FP and FN: new results.

| Total patterns | Well-classified | False positive |
|----------------|-----------------|----------------|
| 2000 | 1870 | 130 |
| **True Positive** | **True Negative** | **False Negatives** |
| 1000 | 870 | 0 |
| **Sensitivity** | **Specificity** | **Accuracy** |
| 1 | 0.915 | 95.75% |

## 7. DISCUSSION

Several trial and error were required to obtain the good results presented in this work. It is shown that the algorithm works so much better using the Deep Learning architecture because the autoencoder is a powerful tool which extract information which maybe is hidden for the BPA only. These results are reflected in the poor stability of the BPA, in the other hand, the Deep Learning architecture has a high stability.

It is expected to do further research in this area using more variability in the configuration of the Deep Learning architecture and incorporate self-tuning techniques like agents and genetic algorithms. Testing the network using different datasets will be an important advance for this kind of researches or even the creation of an own dataset to have the total control of all characteristics of data.

## 8. CONCLUSION

This paper describes a Deep Learning environment where a tuned shallow network and an autoencoder trained with unlabelled big data, collaborate as to produce a reliable a trainable intrusion detection system, operating in access to web situations. The system uses the CICIDS2017 dataset and the autoencoder to produce abstract representation of the activities given into the computer systems. This compressed information is fed to a secondary shallow network that completes the possible attack entry prediction. It is proven that the composed Deep Learning network has a better performance than any other shallow network variation. The resultant functional MATLAB code behaves as a re-trainable, fast, reliable system proved with real data.

### ACKNOWLEDGEMENTS

The authors thankfully acknowledge The Canadian Institute for Cybersecurity (CIC) since they freely provided their research data.

## AUTHORS

**Sergio Hidalgo-Espinoza** is a teaching technician from University of Yachay Tech. He is a graduate of the same university in 2019 as Engineer in Information Technology. His main interests in researching are Information Security, Artificial Neural Networks, High-Performance Computing and Mathematical Modelling.

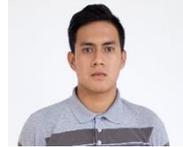

**Kevin Chamorro-Cupuerán** is a teaching technician from University of Yachay Tech. He is a graduate of the same university in 2019 as Mathematician. His main interests in researching are Artificial Neural Networks, Statistics, Mathematical Modelling and Data Mining.

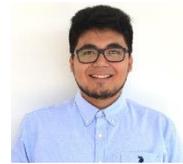

**Professor Dr. Oscar Chang-Tortolero** is an experimented researcher in Artificial Neural Networks. He is actually full-professor at University of Yachay Tech.

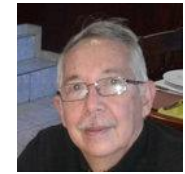